\documentclass[british]{iopart}
\usepackage[T1]{fontenc}
\usepackage[latin1]{inputenc}
\usepackage{graphicx}

\makeatletter

\usepackage{iopams}
\usepackage{setstack}

\usepackage[T1]{fontenc}
\usepackage{ae,aecompl}

\usepackage{babel}
\makeatother

\begin{document}

\title{Time-averaging within the excited state of the nitrogen-vacancy centre
in diamond}

\author{L J Rogers, R L McMurtrie, S Armstrong, M J Sellars and N B Manson}

\address{Laser Physics Center, RSPhysSE, Australian National University, Canberra,
ACT 0200, Australia}

\ead{lachlan.rogers@anu.edu.au}

\begin{abstract}
The emission intensity of diamond samples containing nitrogen-vacancy
centres are measured as a function of magnetic field along a $\langle111\rangle$
direction for various temperatures. At low temperatures the responses
are sample and stress dependent and can be modeled in terms of the
previous understanding of the $^{3}\! E$ excited state fine structure
which is strain dependent. At room temperature the responses are largely
sample and stress independent, and modeling involves invoking a strain
independent excited state with a single zero field splitting of 1.42
GHz. The change in behaviour is attributed to a temperature dependent
averaging process over the components of the excited state orbital
doublet. It decouples orbit and spin and at high temperature the spin
levels become independent of any orbit splitting. Thus the models
can be reconciled and the parameters for low and high temperatures
are shown to be consistent.
\end{abstract}


\noindent{\it Keywords\/}: {\noindent excited state, electronic structure, nitrogen-vacancy center,
diamond}

\maketitle

\section{Introduction}

There is interest in using the $\mbox{NV}^{-}$ centre in diamond
for applications in the area of quantum information processing \cite{Wrachtrup2006},
quantum computing \cite{Jelezko2004a,Jelezko2004b,Dutt2007,Neumann2008},
and magnetometry \cite{Balasubramanian2008,Maze2008}. The applications
rely on the spin in the ground state, and the behavior of this spin
in its environment is largely understood \cite{Maze2008,Takahashi2008,Hanson2008,Jiang2008}.
The excited state is also used, for initialization and for readout,
and an understanding of spin in this state has been obtained from
low temperature excitation measurements \cite{Tamarat2008}. The state
is an orbital doublet, which is usually split by internal strain,
and the excitation measurements examined the fine structure in both
orbital branches. More recently, however, fine structure has also
been determined from room temperature ODMR measurements \cite{Fuchs2008,Neumann2008}
and it is not clear how these values related to those obtained for
the two branches at low temperatures. In this work it is shown that
the room temperature results do not correspond to either branch, but
rather are due to an averaging of both branches.

The excited state fine structure has been obtained from high resolution
optical excitation measurements of single NV centres \cite{Tamarat2008,Batalov2009}
and from two laser hole burning \cite{Reddy1987,Manson1994,Santori2006}.
These techniques are applicable at low temperatures but can not be
extended to higher temperatures. On the other hand, spin information
has also been obtained from optically detected magnetic resonance
(ODMR) of single centres \cite{Fuchs2008,Neumann2009}. In this case
the observations were made at room temperature and the techniques
are not readily extended to low temperature. Instead of these techniques,
we use level crossing of the spin states as indicated by changes of
emission as a function of magnetic field. This approach is applicable
over the entire temperature range from room temperature to helium
temperatures, and does not requrie the same level of specialist equipment.
Room temperature responses have been reported previously \cite{Epstein2005},
and the close relationship between ensemble and single site results
suggested that a study of concentrated crystals could be fruitful.
Ensemble measurements using crystals allow uniaxial stress to be introduced;
and so the studies can involve variation of the temperature, the magnetic
field and the stress. The focus of this investigation was how the
averaging process changes the observation between high and low temperatures.

\section{Emission Variation v Magnetic field: Room Temperature\label{sec:Room-Temperature-Experiments}}

\begin{figure}
\hspace{71pt}\includegraphics[width=8.6cm]{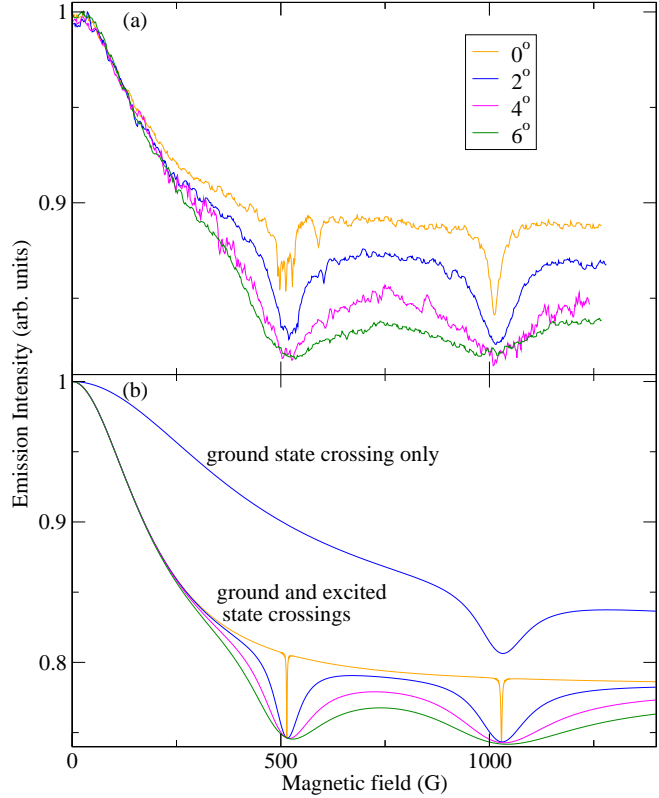}

\caption{(a) Measured magnetic spectra at room temperature for various misalignment
angles as indicated in the common legend. The fine structure on the
510 G feature in the experimental trace for $0^{\circ}$ misalignment
is due to spin cross-relaxation with nearby nitrogen impurities, and
is not of interest here. (b) Calculated magnetic spectra for the same
misalignment angles, simulating an ensemble sample with four different
$\mbox{NV}^{-}$ orientations. The variation in emission intensity
was calculated using rate equations for the 7-level model shown in
Figure \ref{fig:Energy-level-diagram}. The upper trace shows the
effect of spin mixing (spin level avoided-crossing) in the ground
state at 1028 G, and the other traces also include similar spin mixing
in the excited state at 510 G. \label{fig:vs-angle-data-with-model}}

\end{figure}

\begin{figure}
\hspace{71pt}\includegraphics[width=8.6cm]{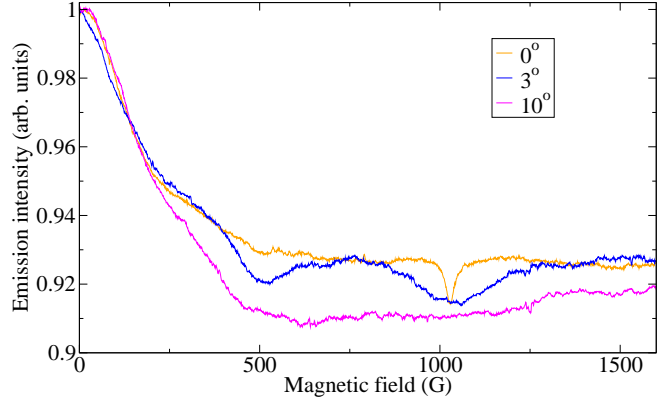}

\caption{Room temperature magnetic spectra for a sample with low concentrations
of interstitial nitrogen. The fine structure at 510 G is absent even
for a well-aligned field. \label{fig:Low-nitrogen-data}}

\end{figure}

A sample (designated L) with an $\mbox{NV}^{-}$ concentration of
$10^{-17}$ ppm was aligned with a $\langle111\rangle$ face parallel
to the magnetic field direction. The sample was excited using a 532
nm wavelength laser and the red emission at right angles detected
using a Si diode. This emission was monitored as a function of magnetic
field and is shown in Figure \ref{fig:vs-angle-data-with-model}(a).
There is about a $10\%$ drop in intensity between 0 gauss and 400
gauss followed by a flatter response, with sharp features at $510\pm5$
gauss and $1028\pm3$ gauss. The sharp dip at 1028 gauss occurs when
there is an avoided crossing of the ground state spin levels for centres
aligned with the magnetic field \cite{Holliday1989,Oort1989,Epstein2005}.
This feature gets narrower with closer alignment of the $\langle111\rangle$
axis with the magnetic field \cite{Martin2000a}, and adjustment was
made to minimize the width. Emission can also be quenched by radio
waves. The minimum frequency (20 MHz) at which this could be achieved
indicated that the transverse field was about three gauss, and hence
the misalignment was less than 0.2 degrees (taken to be 0 degrees).
The emission responses as a function of magnetic field (magnetic spectra)
were then measured for angles of 2, 4 and 6 degrees and these are
also shown in Figure \ref{fig:vs-angle-data-with-model}(a). The feature
of primary interest is the decrease at $510$ gauss, but it is complicated
by cross relaxation with single interstitial nitrogen occurring at
the same field value (see Section \ref{sec:Other-Features}). As can
be seen in Figure \ref{fig:Low-nitrogen-data} the cross relaxation
feature is not present when the sample (N) contains few substitutional
nitrogen impurities. The emission of this sample was low and not convenient
for the more detailed studies.

\begin{figure}
\hspace{71pt}\includegraphics[width=8.6cm]{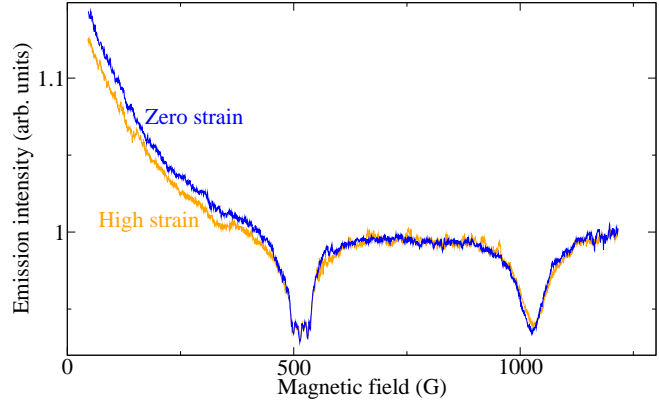}

\caption{Room temperature magnetic spectra (as in Figure \ref{fig:vs-angle-data-with-model}(a))
with and without 0.35 GP uniaxial strain applied perpendicular to
the $\langle111\rangle$ crystal axis with which the magnetic field
was aligned. \label{fig:Room-temperature-strain.}}

\end{figure}

Uniaxial stress of up to 0.5 GP could be applied using a piston arrangement.
It was applied along the $\langle011\rangle$ direction and transverse
to the centres giving rise to 1028 gauss and 510 gauss features. The
results are shown in Figure \ref{fig:Room-temperature-strain.} and
it can be seen that the stress has no affect on the room temperature
response.

\section{Interpretation of Room Temperature Response}

\begin{figure}
\hspace{71pt}\includegraphics[width=4.3cm]{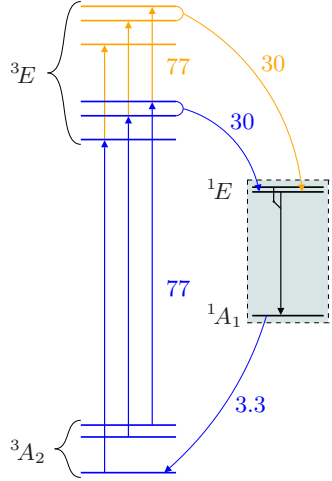}

\caption{Energy level models used for calculations, with transition rate parameters
in units of $10^{6}\mbox{s}^{-1}$ from \cite{Manson2006}. Non spin-conserving
transitions were included in calculations to give realistic spin polarisation
levels, but are left off this diagram for clarity. The structure in
the intermediate singlets is not important here since the lifetime
of the $^{1}\! E^{\prime}$ level is very short \cite{Rogers2008},
and so the system enclosed in the dashed box can be treated as a single
level. The 7-level model includes the three ground spin levels, a
single intermediate level and three spin excited state levels (and
is shown in blue). To describe the low temperature results, it is
necessary to add the three spin levels for the second orbital branch
in the excited state (shown in orange) to give a 10-level model. \label{fig:Energy-level-diagram}}

\end{figure}

The optical transition associated with the nitrogen-vacancy defect
is known to be due to an $^{3}\! A_{2}\leftrightarrow^{3}\! E$ transition
at a site of trigonal symmetry. Neglecting the orbital doublet nature
of the excited state, the dynamics have been described in terms of
a 7 level model; three spin levels in the ground state, three in the
excited state, and an intermediate singlet level \cite{Manson2006}
(The $m_{s}=\pm1$ spin levels in the ground and excited states can
be degenerate and if these are treated as single levels the system
can be considered as a 5 level system). These energy levels are shown
schematically in Figure \ref{fig:Energy-level-diagram}, with transition
rate parameters which have been obtained from lifetime measurements
and other transient changes in the emission \cite{Manson2006}. The
dominant terms are those associated with the optical transitions where
spin projection is conserved, and those associated with decay via
the singlet. The singlet path results in a fraction of the $m_{s}=\pm1$
excited spins relaxing to the ground $m_{s}=0$ without visible emission,
and as a consequence the emission associated with $m_{s}=\pm$1 is
weaker than that associated with $m_{s}=0$. Once the system is cycled
several times the population is predominantly in the $m_{s}=0$ level
(the system is {}``spin polarized'') and the emission is high. Spin
polarization and emission will be reduced whenever the spin states
are mixed, and this can occur with the application of an external
magnetic field. The mixing introduced by a magnetic field is readily
calculated and a rate equation treatment of the 7 levels can be used
to predict the changes in the intensity of the emission as a function
of the magnetic field (Figure \ref{fig:vs-angle-data-with-model}(b)).

The largest changes occur for magnetic fields giving rise to crossing
of the spin levels. For example, the zero field separation between
the $m_{s}=\pm1$ and $m_{s}=0$ spin states in the ground state is
2.88 GHz and so if an axial field of 1028 gauss ($g=2$) is applied
the $m_{s}=0$ and $m_{s}=-1$ become degenerate. Any additional transverse
magnetic field lifts this degeneracy and the basis states become totally
mixed. Spin polarization is no longer maintained and the emission
intensity is reduced. This accounts for why there is a sharp reduction
in the emission intensity when a magnetic field aligned close to a
$\langle111\rangle$ direction is swept in magnitude through 1028
gauss. With misalignment, and thus a larger transverse field, the
spin mixing occurs over a larger range of field and accounts for the
broadening of the features. For ensemble samples the non-axial centres
also contribute to the emission and there is an additional change
in emission as a function of magnetic field. For a field sweep from
0 to 1400 gauss calculation gives a gradual decrease in emission intensity
plus the sharp dip at 1028 gauss. Such emission changes are observed,
but the calculation only allowing for this ground state avoided crossing
does not give good correspondence with the observations as can be
seen by comparing upper trace in Figure \ref{fig:vs-angle-data-with-model}(b)
with experiment in Figure \ref{fig:vs-angle-data-with-model}(a).

In the present work we also consider the effect of a zero field splitting
in the excited state. To account for a level crossing at $510\pm5$
G, the zero field splitting is taken to be $1.43\pm0.01$ GHz (with
$g=2$). The effect on the eigenstates can be calculated in the same
ways as for the ground state. The excited state avoided crossing causes
a decrease at the field value of 510 gauss which again broadens with
misalignment. Also the contribution for the non-axial centres gives
reduced emission at low field values. With this extended rate equation
model the combined effect of the ground and excited state avoided
crossings is found to give excellent agreement with experiment (Figure
\ref{fig:vs-angle-data-with-model}).

It is concluded that the room temperature magnetic field data can
be explained assuming there is an excited state with a single avoided
crossing at 510 gauss. Within experimental uncertainty, this is consistent
with an state having a zero field splitting of 1.42 GHz and $g=2$
as has been observed in the room temperature ODMR measurements \cite{Fuchs2008,Neumann2008}.

\section{Other Features\label{sec:Other-Features}}

The interpretation for well aligned fields must be considered more
carefully as the rate equation model used above is unreliable. In
the model it has been assumed that the optical transitions occur between
the eigenstates of the ground and excited state but time scales can
become too short for the system to relax to its eigenstates. Consider,
for example, the situation immediately after there is decay from the
singlet into the $m_{s}=0$ level in the ground state. Where there
is a ground state avoided crossing, the effective magnetic fields
are all weak (only the weak transverse field is effective) and so
the spin projection precesses slowly. With intense excitation, the
system can be pumped out of the ground state before there is any reorientation
of the spin. The spin polarization and high emission will be maintained.
This is inconsistent with excitation out of the eigenstates which
would conclude that $m_{s}=0$ and $m_{s}=-1$ are equally populated
giving no spin polarization and low emission. Similarly if there is
avoided crossing in the excited state the spontaneous decay can occur
before there is reorientation, and again emission remains high inconsistent
with the result of emission from the eigenstates. Whenever the misalignements
are large the reorientation of the spin projections occurs quickly
compared to the lifetimes, and the time average values gives the same
result as with the eigenstate description. The calculations used above
are valid for misalignments greater than 1 degree, and this is the
range considered by the experiments used in this paper.

As just discussed, when the magnetic field is aligned within 1 degree
the axial centre maintains good spin polarization even at 510 gauss.
However, at this field value the spin polarization can be quenched
due to a cross relaxation process. For example at room temperature
at 510 gauss there is an energy match of the separation of the spins
of the NV$^{-}$ centre and $S=\frac{1}{2}$ centres such as substitutional
nitrogens. Due to the energy match the optically induced spin polarization
can not then be maintained in the presence of spins with reversed
spin population. This cross relaxation including the involvement of
changes of nuclear spin projection has been the subject of several
studies \cite{Holliday1989,Oort1989,Glasbeek1990,Hanson2006,Gaebel2006,Hanson2006a,Jacques2009}.
Such a feature can be seen in the $0^{\circ}$ misalignment trace
on Figure \ref{fig:vs-angle-data-with-model}(a), and it clearly has
structure. The feature is observed whenever there is a level of spin
polarization. This cross relaxation feature is not the subject of
the present study and adds a level of complexity for the interpretation
of the spectra.

\section{Emission Variation v Magnetic field:Low Temperature}

\begin{figure}
\hspace{71pt}\includegraphics[width=8.6cm]{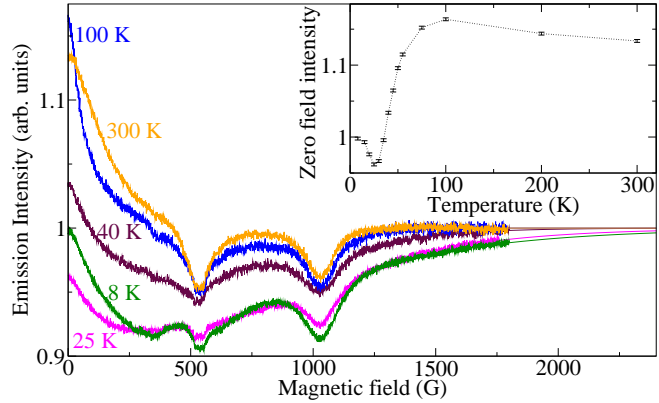}

\caption{Temperature dependance of the magnetic field spectrum. The spectra
have been fitted beyond 1250 G with exponential functions, and normalised
at the high field limit. The fine structure at 514 G is not related
to the excited state, but provides useful confirmation of the field
magnitude. The inset shows the zero field to high field ratio as a
function of temperature. \label{fig:Temperature-dependance}}

\end{figure}

Magnetic field spectra were recorded for temperatures between 8 and
300 K, and the variation for sample L is shown in Figure \ref{fig:Temperature-dependance}.
Due to thermal expansion of the components holding the diamond sample
and thermal variation in absorbtion it was difficult to keep the absolute
magnitude calibrated over a large temperature range. However the emission
is determined by the optical pumping cycle, and at high field will
have a fixed value as the spin eigenstates are (to a good approximation)
determined by the magnetic field and thus are independent of temperature.
The emission responses have, therefore, been normalised to their high
field intensities. An exponential fit to the high field tail of the
measured spectra allowed each trace to be extrapolated to its asymptotic
value.

The 1028 G feature that corresponds to an avoided crossing in the
ground state is approximately constant, indicating that the ground
state properties do not change with temperature. However, the 510
G feature attributed to the excited state changes dramatically. At
room temperature it is centred on 510 G, slightly narrower than the
ground state crossing feature, and retains these properties down to
100 K. Cooling further to 40 K causes this feature to diminish significantly.
By 25 K there is no feature at 510 G, and instead there is a local
minimum in the spectrum at about 250 G which is much broader than
the ground state crossing. Cooling below 25 K caused a double-bump
to appear (Figure \ref{fig:Temperature-dependance}). Also it is noted
that as the sample is cooled the emission intensity at zero-field
varies significantly (inset in Figure \ref{fig:Temperature-dependance}).
It has a minimum at 25 K and a maximum at 100 K. There is a small
reduction in zero-field intensity as the temperature increases from
100 K to room temperature.

\begin{figure}
\hspace{71pt}\includegraphics[width=8.6cm]{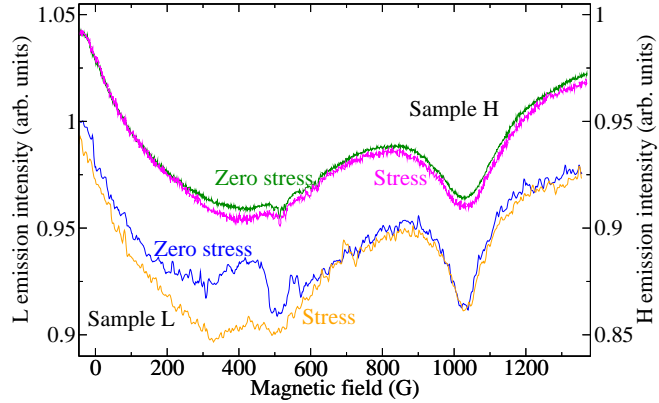}

\caption{Low temperature (15 K) magnetic spectra with and without 0.35 GP uniaxial
stress perpendicular to the $\langle111\rangle$ crystal axis with
which the magnetic field was aligned. The traces for Sample H (higher
concentration) have been translated up for clarity and correspond
to the right-hand vertical axis. \label{fig:Strain-dependence-low-temp}}

\end{figure}

Unlike the effect at room temperature, uniaxial stress was found to
alter the magnetic field spectrum at low temperature. Examples are
shown in Figure \ref{fig:Strain-dependence-low-temp}. Sample L exhibits
significant changes with stress. The doublet is changed to single
broad feature (plus the cross relaxation feature). The second sample
H contained a higher concentration of $\mbox{NV}^{-}$ defects and,
although barely changed by external stress, already had the characteristics
of the first sample with stress applied.

\section{Interpretation of Low Temperature Response}

\begin{figure}
\hspace{71pt}\includegraphics[width=8.6cm]{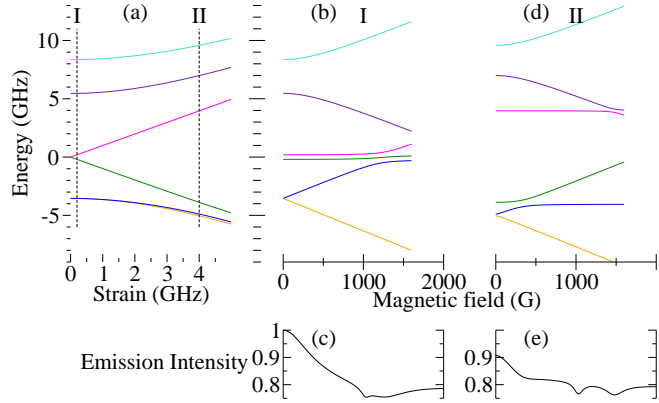}

\caption{(a) Excited state energy level splittings as a function of strain.
(b) Splitting of energy levels for a centre at strain I (0.2 GHz)
with a magnetic field applied at $2^{\circ}$ to the $\mbox{NV}^{-}$
axis, and (c) the magnetic spectra calculated for this strain from
the 10-level model in Figure \ref{fig:Energy-level-diagram}. (d)
and (e) show the corresponding responses to an identical magnetic
field when the centre is at strain II (4 GHz). The magnetic spectra
in (c) and (e) are shown in more detail in Figure \ref{fig:Calculated-low-temp-mag-spec}(a),
but are included here to indicate their connection with the spin level
avoided crossings. \label{fig:Energy-level-splittings}}

\end{figure}

At low temperatures fine structure attributed to the $^{3}\! E$ state
has been observed in holeburning \cite{Reddy1987,Redman1992,Manson1994,Santori2006},
excitation \cite{Tamarat2008,Batalov2009} and photon echo \cite{Rand1994,Lenef1996a}
experiments. The Hamiltonian associated with this structure is given
by

\begin{equation}
\mathcal{H}=\mathcal{H}_{so}+\mathcal{H}_{ss}+\mathcal{H}_{str}+\mathcal{H}_{z}\label{eq:full-hamiltonian}\end{equation}

where $\mathcal{H}_{so}$ is spin-orbit, $\mathcal{H}_{ss}$ spin-spin,
$\mathcal{H}_{str}$ strain and $\mathcal{H}_{z}$ the Zeeman interaction.
The diagonal component of spin-orbit $\mathcal{H}_{so}=\lambda_{z}\mathbf{L}_{z}\mathbf{S}_{z}$
splits the state into three equally separated doublets. One doublet
($E$ irreducible representation) is not displaced and associated
with the $S_{z}$ states ($m_{s}=0$). The state where spin and orbital
projections are parallel (also $E$ irreducible representation) is
displaced down in energy by $\lambda_{z}$ and the antiparallel alignments
($A_{1}$ and $A_{2}$ states) are displaced up in energy by the same
magnitude. The off-diagonal spin-orbit $\lambda_{x,y}\left(\mathbf{L}_{x}\mathbf{S}_{x}+\mathbf{L}_{y}\mathbf{S}_{y}\right)$
has only a minor effect as its value is known from previous work to
be small ($\lambda_{x,y}=0.2\mbox{ GHz}$) \cite{Tamarat2008}. Spin-orbit
can mix $^{3}\! E$ states with singlets and is important for population
decay, but has negligible effect on the energy levels and is not included
here. Spin-spin $\mathcal{H}_{ss}$ has two contributions. There is
a normal spin-spin term $D_{es}\left(\mathbf{S}_{z}^{2}-\frac{2}{3}\right)$
which shifts the $m_{s}=\pm1$ states with respect to $m_{s}=0$ states
by $D_{es}$. The second spin-spin contribution is associated with
spins on the two orbital components of the degenerate state \cite{Lenef1996}.
It only affects $A_{1}$ and $A_{2}$ states and we assign the magnitude
of this splitting as $\Delta$. The energy levels in perfect C$_{3v}$
symmetry are therefore shown on the left of Figure \ref{fig:Energy-level-splittings}
\cite{Tamarat2008,Manson2007}. The energies are determined by four
parameters and the value of the parameters have been obtained from
high resolution excitation traces of single centres. The most reliable
parameters from recent measurements \cite{Batalov2009} are $\lambda_{z}=5.5\mbox{ GHz}$,
$D_{es}=1.42\mbox{ GHz}$, $\Delta=3.1\mbox{ GHz}$, and $\lambda_{x,y}=0.2\mbox{ GHz}$.

The splitting of the excited state energy levels with strain, as shown
in Figure \ref{fig:Energy-level-splittings}(a), can be calculated
by diagonalising the energy matrix

\[
\fl\mathcal{H}=\left[\begin{array}{cccccc}
\begin{array}{c}
\lambda_{z}\\
+\frac{1}{3}D_{es}\\
+\Delta\end{array} & 0 & 0 & 0 & \delta_{y} & \delta_{x}\\
0 & \begin{array}{c}
\lambda_{z}\\
+\frac{1}{3}D_{es}\\
-\Delta\end{array} & 0 & 0 & \delta_{x} & \delta_{y}\\
0 & 0 & \begin{array}{c}
-\frac{2}{3}D_{es}\\
-\delta_{y}\end{array} & \delta_{y} & i\lambda_{x,y} & 0\\
0 & 0 & \delta_{y} & \begin{array}{c}
-\frac{2}{3}D_{es}\\
+\delta_{y}\end{array} & 0 & i\lambda_{x,y}\\
\delta_{y} & \delta_{x} & -i\lambda_{x,y} & 0 & \begin{array}{c}
-\lambda_{z}\\
+\frac{1}{3}D_{es}\end{array} & 0\\
\delta_{x} & \delta_{y} & 0 & -i\lambda_{x,y} & 0 & \begin{array}{c}
-\lambda_{z}\\
+\frac{1}{3}D_{es}\end{array}\end{array}\right]\]
where $\delta_{x}$ and $\delta_{y}$ are strain parameters.

Strain perpendicular to the trigonal axis lowers the symmetry and
$\mathcal{H}_{str}$ has the effect of splitting the $^{3}\! E$ into
two orbital branches each with three spin levels (Figure \ref{fig:Energy-level-splittings}(a))
\cite{Tamarat2008,Manson2007}. In the lower branch spin-orbit and
spin-spin have opposite signs. At zero strain spin-orbit is larger
resulting in the $m_{s}=\pm1$ states being lowest. However, spin-orbit
splitting is reduced with increasing strain and the separation of
the $m_{s}=\pm1$ and $m_{s}=0$ levels is reduced. The states can
cross, although the crossing is changed to an avoided crossing by
off-diagonal spin-orbit interaction. The distance of closest approach
of these levels enables the strength of the interaction to be determined
and a value of $\lambda_{x,y}=0.2$ has been obtained in this way
\cite{Tamarat2008}. In the upper branch, spin-orbit and spin-spin
have the same sign and the states are always well separated such that
there is little mixing \cite{Tamarat2008,Manson2007}. 

\begin{figure}
\hspace{71pt}\includegraphics[width=8.6cm]{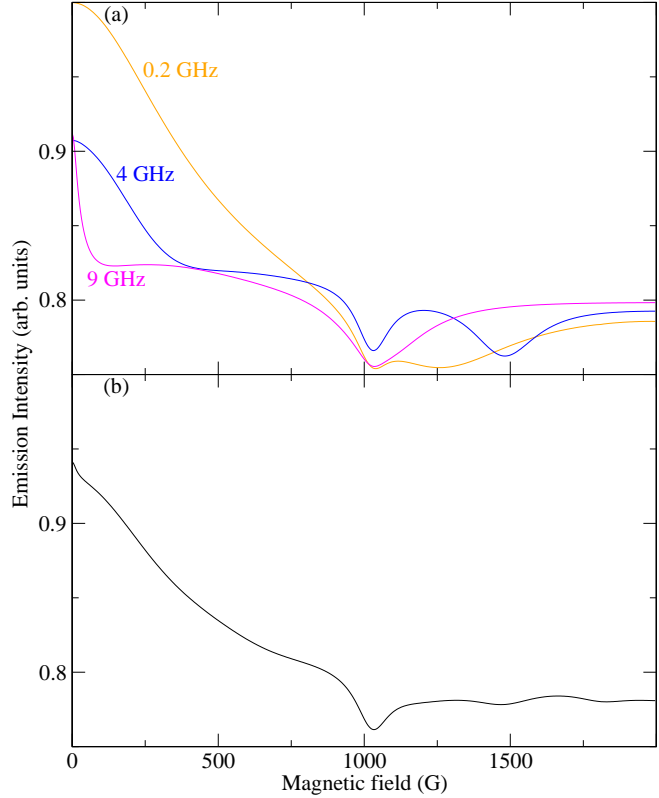}

\caption{Calculated magnetic spectra, with a field misalignment of $2^{\circ}$,
for (a) strain splittings of 0.2, 4 and 9 GHz; and (b) an ensemble
with strain distribution. The strain distribution in (b) consisted
of 0, 0.2, 2, 4, 7, 9 GHz with corresponding weights of 1, 3, 5, 5,
3, 1. The off-axis $\mbox{NV}^{-}$ centres are included in these
calculations, simulating an ensemble crystal sample. \label{fig:Calculated-low-temp-mag-spec}}

\end{figure}

For low fields ($<2500$ gauss) the Zeeman interaction is dominated
by $2\mathbf{S}\beta\mathbf{B}$ and the splittings are illustrated
in Figure \ref{fig:Energy-level-splittings}(b) and (d) for centres
with various values of strain. In the case where strain is sufficient
to split the $^{3}$E state into two triplets (II) each triplet is
split analogous to that of an orbital singlet. The important features
are the avoided crossings of the $m_{s}=0$ and $m_{s}=-1$ spin levels.
There is one associated with each orbital branch. Rate equations for
the 10 levels (Figure \ref{fig:Energy-level-diagram}) are used to
calculate the changes of emission as a function of field. Decrease
in emission is obtained at field values corresponding to the two excited
state avoided crossings plus the one in the ground. As can be seen
from the examples magnetic field value at which the excited state
dips occur varies with strain and can occur at zero field (Figure
\ref{fig:Calculated-low-temp-mag-spec}(a)).

For an ensemble the emission will have contribution from centres with
a range strain environments, and an example using six strain values
is shown in Figure \ref{fig:Calculated-low-temp-mag-spec}(b). Reliable
calculation would require knowledge of the distribution, but this
is not available. However, we can comment on how certain characteristics
arise. A contribution from centres with mixing at zero field will
depress the emission at zero field and this has been observed in the
experiment (Figure \ref{fig:Temperature-dependance}). Also as the
avoided crossing in the lower branch always occurs at low field values
($<500$ gauss) it is understandable why the emission in the 200 -
500 gauss range is low. Conversely avoided crossing in the upper branch
occur at higher fields and accounts for the slow increase in emission
intensity as the field is extended above 2000 gauss.

The conclusion is that the low temperature responses are consistent
with that expected for the known fine structure of the NV centre and
its variation with strain. A variation from centre to centre and sample
to sample is understandable in this model, but this is totally different
to the behavior at room temperature.

\section{MCD of Zero-Phonon Line}

An axial magnetic field splits an orbital $E$ state into two components
$E_{+}$ and $E_{-}$ and these components give allowed transitions
to an orbital $A$ state in right and left circular polarized light
respectively. Measurements of the electronic transition can be made
using modulation techniques, and by using the first moment of the
MCD the Zeeman splitting can be determined even in the presence of
inhomogeneous broadening \cite{Shepherd1968}. The splitting allows
the orbital angular momentum to be determined and for an $A\leftrightarrow E$
would be temperature independent The situation here is slightly more
complicated as the transition is $^{3}\! A\leftrightarrow^{3}\! E$
rather than an $A\leftrightarrow E$. For this case the electronic
Zeeman term leads to three sets of $A\leftrightarrow E$ transitions,
one for each spin projection ($m_{s}=0,1\mbox{ and }-1$) and the
MCD of each has a slightly different magnitude \cite{Davis1968}.
Spin-orbit interaction has the effect of increasing the MCD for $m_{s}=-1$
and reducing it for $m_{s}=+1$. The different magnitudes leads to
a temperature dependence of the MCD. However as spin orbit is only
5.24 GHz the change of magnitude is only 5 $\%$ and much smaller
than considered previously \cite{Reddy1987}.

\begin{figure}
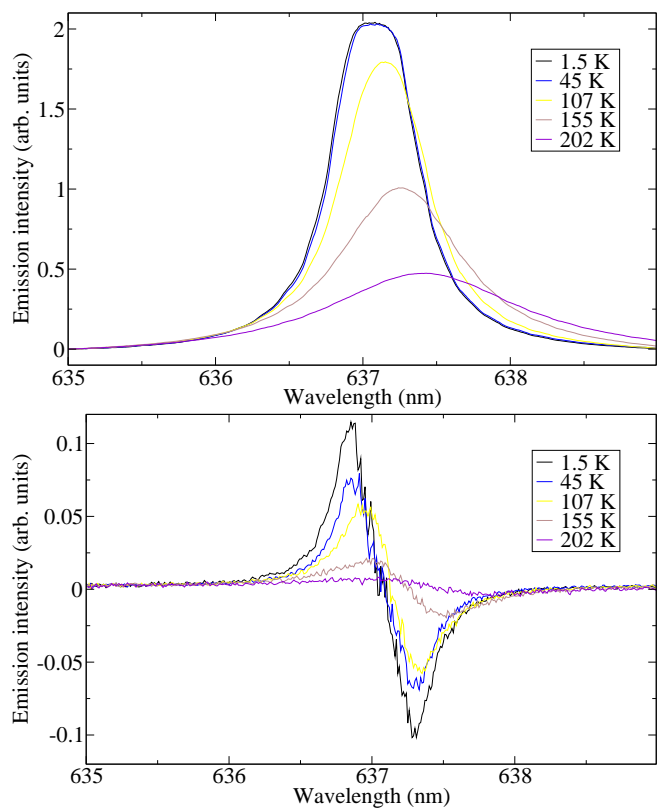

\hspace{71pt}\includegraphics[width=8.6cm]{mcd_traces_absorption}

\hspace{71pt}\includegraphics[width=8.6cm]{mcd_traces_diff}

\caption{Absorption and MCD response for various temperatures. \label{fig:Absorption-and-MCD-with-temp}}

\end{figure}

\begin{figure}
\hspace{71pt}\includegraphics[width=8.6cm]{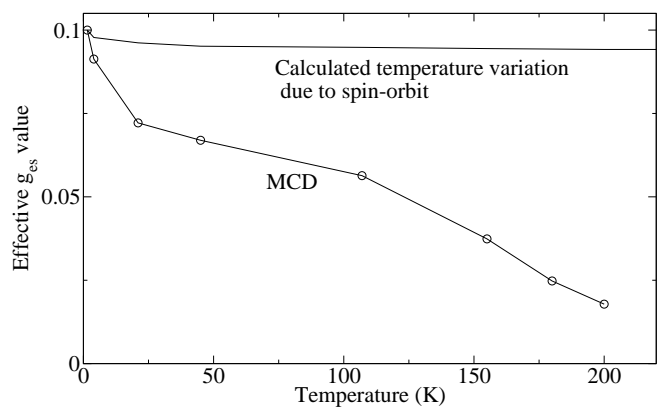}

\caption{Effective $g_{es}$ value as a function of temperature, obtained from
the first moment of the MCD response. \label{fig:MCD-moment-vs-temp}}

\end{figure}

The MCD and absorption of the zero-phonon line at 637 nm was measured
for the $10^{18}$ ppm concentration sample H at various temperatures
using a field of 5 Tesla (\ref{fig:Absorption-and-MCD-with-temp}).
The variation in the first moment normalized to the strength of the
absorption is shown in the insert (Figure \ref{fig:MCD-moment-vs-temp}).
The magnitude of the signal indicates a splitting of 7 GHz consistent
with a $g=0.1$ as obtained previously \cite{Reddy1987}. What is
significant however, is that the signal drops with increasing temperature
and is continuing to drop at 200 K. The size of the decrease is much
larger the 5 per cent that could be attributed to spin-orbit interaction.
The interpretation is that the effective angular momentum is high
at low temperature but approaches zero at room temperature. This variation
will be discussed below.

\section{Averaging Process With Temperature}

The change of behavior with temperature is attributed to an averaging
process and will be discussed here.

It is well known that increasing temperature gives rise to a variation
of the homogeneous line width of optical transitions. Changes of width
involve the transitions combining with the absorption and emission
of lattice vibrations of slightly different energies. The vibrations
also induce transfer between electronic states and give the standard
Boltzman distribution in the population of adjacent states. By considering
specific optical line widths we are able to estimate the time scale
of such processes. At helium temperatures a $^{3}\! A_{2}\leftrightarrow^{3}\! E$
line width of 13 MHz has been observed \cite{Tamarat2006a}and is
determined by the radiative lifetime of 12 ns. The lifetime does not
change with temperature and at higher temperature the larger widths
must reflect lifetimes of the coupled electron-vibrational system.
For example at intermediate temperatures holeburning gives an estimate
of homogeneous line widths and \cite{Nisida1990} indicate values
of the order of 20 GHz at 60 K. Above 180 K the transitions are homogeneously
broadened and can be measured by conventional absorption or emission
but give rates of 1000 GHz \cite{Davies1974}.

Vibrations interact with electronic orbit but not with spin. Therefore
the fast rates will involve rapid interchange between orbital components
of a degenerate state without changing the spin projection. This is
what occurs here, and the fast orbital change has significant effect
on observations. In large magnetic fields the interchange will be
between states, $E_{+},E_{-}$, with positive and negative angular
momentum. Rapid variation will result in an average state with less
or no orbital angular momentum. Thus this type of averaging will reduce
the effective orbital angular momentum and account for the observed
reduction of the MCD.

For strain the interchange will be between orbital component from
$E_{x}$ to $E_{y}$ with the $S_{z},S_{x},S_{y}$ spin projections
remaining unaltered. Although the projections stay the same the energies
change (see Figure \ref{fig:Energy-level-splittings}). Thus normally
the ODMR frequency will be a few GHz's different for the two orbital
branches. However, rapid change of the orbital branch at a rate of
many 10's GHz will alter the situation and the ODMR measurement will
give the average frequency rather than two separate frequencies. (This
is similar to conditions in liquid state NMR.) The averaging effectively
quenches the contribution of spin-orbit interaction $\lambda_{z}$.
Also as $S_{x}$ spin coupled to $E_{x}$ and $E_{y}$ is distributed
between the $A_{1}$ and $A_{2}$ states the averaging quenches the
effect of this splitting $\Delta$. The remaining term in the Hamiltonian
$D_{es}\left(\mathbf{S}_{z}^{2}-\frac{2}{3}\right)$ determines the
average $S_{z}\leftrightarrow S_{x}$ (and $S_{z}\leftrightarrow S_{y}$)
separation and the ODMR frequency. At high strain the spin-spin would
include a strain dependent term $E_{es}\left(\mathbf{S}_{x}^{2}-\mathbf{S}_{y}^{2}\right)$,
and the ODMR will give two excited state resonances at $D_{es}\pm E_{es}$.

The above averaging process results in an 'average' state with a zero
field splitting determined by the spin-spin interaction $D_{es}$
(low strain situation). In order to obtain a fit for recent low temperature
high resolution excitation measurements a zero field splitting of
1.42 GHz is required, and this is confirmed from measurements of the
separation of the averaged energy of $S_{z},S_{x},S_{y}$ levels \cite{Batalov2009}.
Thus there is consistency between the low and high temperature frequencies,
and the averaging accounts for the difference between high and low
temperature behaviours.

The averaging is not associated with a static Jahn-Teller effect.
For modest orbital strain splitting the averaging can result in the
appearance of higher symmetry, however the splitting varies over a
large range from zero to 1000's GHz. Clearly if there is zero strain
splitting there is no Jahn-Teller distortion. Also, with very large
orbital splittings it is observed that the ODMR signal is split and
therefore the system still has low symmetry. In the case of Jahn-Teller
the averaging would result in higher symmetry.

\section{Conclusion}

From this work it is concluded that the spin and optical properties
of the $\mbox{NV}^{-}$ centre are much simpler at room temperature
than has previously been anticipated. Thermal vibrations average the
orbital components of the excited state and quench all orbital contribution
to the relative energy of the spin levels. In effect, the excited
state has the properties of an orbital singlet. Spin level energies
are determined only by spin-spin interaction, just as they are in
the orbital singlet ground state. This state averaging process is
most likely a general phenomenon when thermal transition rates between
orbital branches are large compared to the spin-orbit interaction.
The result is an effective decoupling of spin and orbit. In the $\mbox{NV}^{-}$
system, this averaging has a prominent effect due to the very small
spin-orbit interaction. At low temperature the thermal transition
rates are slowed and spin-orbit interaction does become significant,
and so it is necessary to consider both orbital branches of the excited
state.

This averaging has implications for many earlier studies of the $\mbox{NV}^{-}$
centre. For example, it accounts for why previously observed ODMR
signals can be explained in terms of a spin triplet associated with
a non-degenerate electronic state \cite{Fuchs2008,Neumann2009}. It
also justifies our use of a 7-level model both here and in previous
publications \cite{Manson2006} to explain the optical pumping cycle
of the $\mbox{NV}^{-}$ centre at room temperature.

At room temperature there is little variation in optical properties
from site to site, which could make it simpler to develop applications.
The averaging eliminates the need to seek centres with optimum properties,
and justifies determining many properties from ensemble measurements.
Counterintuitively, stronger spin polarization is obtained at room
temperature than at low temperature, and this is likely to be of considerable
interest for the development of quantum devices.

\ack{}{The authors would like to acknowledge fruitful discussions with Vincent
Jacques, Fedor Jelezko and Jörg Wractrup (Universität Stuttgart).
We would also like to thank Elmars Kraus (Australian National University)
for undertaking the MCD measurements. This work has been supported
by Australian Research Council. }

\section*{References}

\bibliographystyle{unsrt}
\bibliography{/home/lachlan/phd/reference/resources}

\end{document}